# Advancing Antiferromagnetic Nitrides via Metal Alloy Nitridation

*Qianying Wang, Zexu He, Lele Zhang, Qian Li, Haitao Hong, Ting Cui, Dongke Rong, Songhee Choi, Qiao Jin, Chen Ge, Can Wang, Qinghua Zhang, Liang Cheng, Jingbo Qi, Kui-juan Jin,\* Gang-Qin Liu\* and Er-Jia Guo\**

Q. Y. Wang, Z. He, H. T. Hong, T. Cui, D. K. Rong, S. Choi, C. Ge, C. Wang, Q. H. Zhang, K. J. Jin, G. Q. Liu, and E. J. Guo
Beijing National Laboratory for Condensed Matter Physics and Institute of Physics, Chinese Academy of Sciences, Beijing 100190, China
E-mails: kjjin@iphy.ac.cn, gqliu@iphy.ac.cn and ejguo@iphy.ac.cn

Q. Y. Wang, H. T. Hong, T. Cui, D. K. Rong, C. Ge, C. Wang, Q. H. Zhang, K. J. Jin, G. Q. Liu, and E. J. Guo
Department of Physics & Center of Materials Science and Optoelectronics Engineering, University of Chinese Academy of Sciences, Beijing 100049, China

L. Zhang, L. Cheng, J. Qi
State Key Laboratory of Electronic Thin Films and Integrated Devices, University of Electronic Science and Technology of China, Chengdu 611731, China

Q. Li
National Synchrotron Radiation Laboratory, University of Science and Technology of China, Hefei 230029, China

**Abstract**

Nitride materials, valued for their structural stability and exceptional physical properties, have garnered significant interest in both fundamental research and technological applications. The fabrication of high-quality nitride thin films is essential for advancing their use in microelectronics and spintronics. Yet, achieving single-crystal nitride thin films with excellent structural integrity remains a challenge. Here, we introduce a straightforward yet innovative metallic alloy nitridation technique for the synthesis of stable single-crystal nitride thin films. By subjecting metal alloy thin films to a controlled nitridation process, nitrogen atoms integrate into the lattice, driving structural transformations while preserving high epitaxial quality. Combining nanoscale magnetic imaging with a diamond nitrogen-vacancy (NV) probe, X-ray magnetic linear dichroism, and comprehensive transport measurements, we confirm that the nitridated films exhibit a robust antiferromagnetic character with a zero net magnetic moment. This work not only provides a refined and reproducible strategy for the fabrication of nitride thin films but also lays a robust foundation for exploring their burgeoning device applications.

**Introduction**

Spintronics is an emerging field in condensed matter physics that integrates processing, storage, sensing, and logic functions into a unified platform. Compared to conventional electronic devices, spintronic devices offer significant advantages in data processing speed, non-volatility, and energy efficiency.[1-4] However, to compete with silicon-based semiconductor technology, continued miniaturization and innovation in spintronic device design are essential.[5] A key challenge in spintronic devices based on ferromagnetic (FM) materials is the presence of stray fields, which become increasingly problematic at reduced dimensions. These fields can induce edge domains and unwanted crosstalk between junctions, limiting the scalability and operational reliability of FM-based spintronic architectures. In contrast, antiferromagnetic (AFM) materials inherently lack stray fields, making them ideal candidates for high-density spintronic applications. Furthermore, AFM materials exhibit ultrafast spin dynamics, exceptional robustness against external perturbations, and enhanced thermal stability, offering significant advantages over their FM counterparts. Their insensitivity to external magnetic fields and potential for terahertz-frequency operation position AFM spintronics as a transformative approach for next-generation information technology.[6-9]

Compared to traditional oxide- and metal-based antiferromagnets, nitride-based antiferromagnetic materials[10-16] exhibit the enhanced spin current transport efficiency. Simultaneously, the incorporation of highly electronegative nitrogen atoms into interstitial sites of metallic lattices strengthens bonding interactions, enhancing mechanical and thermal stability, as well as the electronic tunability.[17-18] These characteristics provide more effective spin polarization and control, positioning them as promising candidates for next-generation spintronic devices. The fabrication of nitride thin films is conventionally achieved using techniques such as magnetron sputtering, molecular beam epitaxy (MBE), pulsed laser deposition (PLD), and chemical vapor deposition (CVD). For example, Nan *et al.* utilized DC reactive planar magnetron sputtering to produce high-quality $Mn_3GaN$ films with atomically sharp interfaces.[19-20] Jin *et al.* demonstrated the fabrication of stoichiometric CrN ultrathin films with high crystallinity using plasma-assisted PLD.[12-13] Azizi *et al.* successfully synthesized large-area and high-quality hexagonal boron nitride (*h*-BN) films via the chemical vapor deposition (CVD) method.[21] Despite their advantages, these techniques present notable challenges. Magnetron sputtering, which relies on chemical reactions between a metal target and a nitrogen-argon gas mixture, often struggles to achieve uniform chemical composition and high crystallinity in nitride films. While MBE can produce high-quality films, it suffers from a slow deposition rate, operational complexity, difficulties in replacing evaporation sources, and limited reproducibility. PLD, although promising, is constrained by challenges in fabricating high-quality nitride targets. Similarly, CVD, despite its ability to produce uniform and conformal films over large areas, faces significant limitations. CVD's elevated processing temperatures induce thermal stress and defects, while the use of toxic precursors presents safety and scalability challenges.[22-24] Moreover, achieving precise control over stoichiometry and crystallinity in multicomponent nitride films remains a significant challenge. As a result, achieving precise control over nitrogen content, maintaining high crystallinity, and developing complex multicomponent nitride structures remain critical challenges in the field.

In this work, we introduce a versatile and effective synthesis route for single-crystal nitride thin films by employing high-temperature ammoniation of alloy thin films prepared via PLD. This approach enables precise nitrogen incorporation, overcoming challenges associated with conventional nitride synthesis techniques while ensuring high crystallinity and uniform chemical composition. The incorporation of nitrogen atoms induces a distinct magnetic phase transition. Before nitridation, the alloy films exhibit a ferromagnetic state, whereas after annealing, they develop a net-zero magnetic moment, signifying the suppression of ferromagnetism. Additionally, the orbital occupancy is modulated by an external magnetic field, revealing emergent antiferromagnetic characteristics.

**Results and discussions**

**Synthesis and structural characterizations of Mn$_3$GaN epitaxy films**

Manganese-based nitride films have previously been synthesized using DC reactive (or RF) magnetron sputtering and nitrogen plasma-assisted molecular beam epitaxy (MBE).[25-27] Experimental evidence reveals that the structural and magnetic properties of these films are highly sensitive to nitrogen stoichiometry, often leading to the formation of secondary phases under varying synthesis conditions, such as temperature and atmospheric composition.[14, 28-30] These secondary phases hinder further investigation into the properties of manganese-based nitride films. Here we introduce a novel method for fabricating single-phase manganese-based nitride thin films, as illustrated in Figure 1a. Initially, stoichiometric alloy films were deposited on the substrates at 500°C under a base pressure of $1 \times 10^{-8}$ Torr. The film thickness was precisely controlled by monitoring the number of laser pulses. Subsequently, the films underwent rapid thermal processing (RTP) annealing, a critical step for facilitating nitrogen incorporation into the metallic lattice and promoting the formation of high-quality single-crystal films. The annealing process involved heating the films to 700°C in a pure ammonia atmosphere for 5 minutes, followed by an hour annealing period and gradual cooling to room temperature.

Mn$_3$Ga alloy films have attracted considerable attention due to their diverse and tunable structural and magnetic phases. Among them, the D0$_{19}$ hexagonal phase is the most thermodynamically stable, characterized by a triangular antiferromagnetic spin order in the kagomé plane.[31-32] Another significant phase is the D0$_{22}$ tetragonal structure (Figure 1b), belonging to the space group I4/mmm. In this tetragonal phase, MnI and MnII atoms occupy the 2b (0, 0, 1/2) and 4d (0, 1/2, 1/4) Wyckoff positions, respectively, contributing to ferromagnetic ordering through their antiparallel spin alignment.[33-35] The nitridation of manganese leads to various phases, heavily dependent on the nitrogen content. For instance, binary manganese nitrides transition through $\theta$-MnN, $\eta$-Mn$_3$N$_2$, $\zeta$-Mn$_5$N$_2$ (-Mn$_2$N, -Mn$_2$N$_{0.86}$) and $\varepsilon$-Mn$_4$N as the nitrogen content decreases. The $\theta$ and $\eta$ phases exhibit a NaCl-type face-centered tetragonal structure at room temperature (Figure 1c), while the $\zeta$ phase adopts a hexagonal structure, both phases demonstrated antiferromagnetic behavior. At the lowest N content (Mn:N ratio of 4:1), the ε phase manifests a ferromagnetic face-centered cubic structure. The substitution of Mn with other metal atoms in binary manganese-based nitride lattices modulates the structure and physical properties of multinary manganese-based nitrides, depending on the doping concentration.[36-38]

Mn$_3$Ga alloy films were fabricated at different growth temperatures with a thickness of approximately 45 nm. These samples were annealed for one hour in an ammonia atmosphere at 700 °C, resulting in a single peak near the substrate peak (Figure S1, Supporting Information). The structural evolution of the samples before and after annealing, represented by the D0$_{22}$ structure, is illustrated in Figures 1b and 1c. To further investigate the crystallographic symmetry, second-harmonic generation (SHG) experiments were performed on Mn$_3$Ga and Mn$_{1-x}$Ga$_x$N films. Figures 2a and 2b show the SHG signals for Mn$_3$Ga and Mn$_{1-x}$Ga$_x$N samples, with emitted light polarization parallel to the incident light polarization ($I_{2\omega}$). The experimental results were fitted by an equation with linear combination of $\cos(2\theta+\phi_2)$, $\cos(4\theta+\phi_4)$, and $\cos(6\theta+\phi_6)$, revealing that signals of Mn$_3$Ga alloy films exhibit both twofold and sixfold symmetries, while those of Mn$_{1-x}$Ga$_x$N films display twofold and fourfold symmetries.

X-ray diffraction (XRD) measurements were performed to reveal the structural evolution after nitridation. Figure 2c shows the XRD *θ-2θ* scans for Mn$_3$Ga film grown at 500 °C on MgO substrate. The Mn$_3$Ga alloy films consist of D0$_{19}$ and D0$_{22}$ phase. After annealing in ammonia, the sample exhibits a single crystalline peak at 43.53°, corresponding to an out-of-plane lattice constant of 4.15 Å. This value aligns well with the reported lattice parameter *c* for θ-MnN, which varies between 4.12 and 4.189 Å depending on the nitrogen content.[16, 28, 39] The inset of Figure 2c presents a narrow rocking curve at Mn$_{1-x}$Ga$_x$N 002 reflection with a full width at half maximum (FWHM) of ≈ 0.06°, indicating high crystallinity. XRD reciprocal space map (RSM) further confirms the coherent growth of Mn$_{1-x}$Ga$_x$N films on MgO substrates, as depicted in Figure 2d. Figure 2e shows a representative atomic-resolved cross-sectional high-angle annular dark-field (HAADF) image of Mn$_{1-x}$Ga$_x$N samples obtained via scanning transmission electron microscopy (STEM). The image confirms the high crystallinity of Mn$_{1-x}$Ga$_x$N single layers, with no obvious defects in the sample. The HAADF-STEM image confirms the coherent growth of the Mn$_{1-x}$Ga$_x$N film on the substrates with an atomically sharp interface. Notably, no discernible brightness difference between Mn and Ga was observed, precluding the analysis of their atomic configurations in the lattice using STEM images. This observation can be attributed to the disordered arrangement of manganese nitride and gallium nitride in the samples after annealing, specifically the random substitution of Mn by Ga within the lattice. The high-magnified STEM image, indicated as the red box in Figure 2e, shows the positions of Mn/Ga atom columns, matching well with the atomic structure of single-phase θ-MnN.[26] Due to its low atomic number (*Z*), nitrogen (N) is not clearly visible in these STEM images. The combined XRD and STEM analyses indicate that the incorporation of nitrogen atoms transforms the film structure from a polycrystalline mixture of hexagonal and tetragonal phases into a single-phase rock-salt crystal structure with high crystallinity. This structural evolution is consistent with the previously observation from SHG.

**Electronic states of high-quality $Mn_{1-x}Ga_xN$ films**

To further elucidate the effect of ammonia annealing treatment on the intrinsic electronic states of the samples, we performed element-specific X-ray absorption spectroscopy (XAS) measurements on $Mn_{1-x}Ga_xN$ at room temperature. Figure 3a illustrates the XAS spectra at the N *K*-edge for the $Mn_{1-x}Ga_xN$ films. Within the gray-shaded region, three distinct peaks are observed at 400.8 eV, 403 eV, and 405.2 eV, corresponding to the N *K*-edges of GaN, as previously reported (solid black line).[40] Outside this region, two additional peaks align with the reported peak positions of $\theta$-$Mn_6N_{5+y}$ (dashed line).[41] These results indicates stable and sufficient chemical bonding between N and Mn/Ga in the annealed samples, suggesting significant structural transformation. XPS analysis of $Mn_{1-x}Ga_xN$ film was carried out by investigating Mn 2*p*, N 1*s* and Ga 3*d* core-level spectra (Figure S2, Supporting Information). The binding energy was carefully calibrated by taking C 1*s* peak at 284.6 eV as a reference. Peak deconvolution revealed a Mn:Ga atomic ratio of 2.9:1, consistent with the stoichiometric ratio of the $Mn_3Ga$ alloy target, corresponding to an *x* value of approximately 0.25 in $Mn_{1-x}Ga_xN$. A single symmetric peak centered at 396.8 eV, corresponding to the N 1*s* state, confirms the presence of nitrogen in the metal nitride. Additionally, the Mn 2*p* peaks at 641.3 eV and 652.8 eV, along with the Ga 3*d* peak at 19.75 eV, indicate strong hybridization between Mn, Ga, and N, corroborating the XAS findings. To probe the electronic properties further, X-ray linear dichroism (XLD) measurements were employed to investigate the unoccupied states (holes) of the $d_{x^2-y^2}$ and $d_{3z^2-r^2}$ orbitals. As illustrated in Figures 3b-3d, the linearly polarized X-rays were incident on the sample surface at angles of 0° and 60° relative to the surface normal. Figures 3e-3g display XAS and XLD spectra at Mn $L$-edges for annealed sample under varying out-of-plane magnetic fields. In $Mn_{1-x}Ga_xN$ films without an external magnetic field, the slight discrepancy between the peak energies of $I_{0°}$ and $I_{60°}$ suggests a preferential occupancy of the $d_{3z^2-r^2}$ orbitals by electrons, leading to a positive XLD value. Upon the application of an external magnetic field perpendicular to the surface of the sample, the $Mn_{1-x}Ga_xN$ films exhibit a negative XLD signal, indicating a shift in electron occupancy to the $d_{x^2-y^2}$ orbitals. Additionally, magnetic characterization was performed to investigate the magnetic properties of the sample. The M-H and M-T curves (Figure S3, Supporting Information) reveal that the $Mn_{1-x}Ga_xN$ film exhibits negligible magnetic response. Previous studies have shown that θ-MnN adopts an antiferromagnetic structure.[42-44] Therefore, we attribute these XLD results to the influence of the antiferromagnetic structure, demonstrating that the orbital occupation state in the $Mn_{1-x}Ga_xN$ films is governed by the applied magnetic field.

**Magnetic transitions probed by diamond NV-based magnetometry**

In addition to the detailed analysis of structure and electronic states, we further investigated the magnetic properties of $Mn_{1-x}Ga_xN$ films. We compared the magnetization of both $Mn_3Ga$ alloy and $Mn_{1-x}Ga_xN$ films. To characterize the magnetic variations induced by annealing, noninvasive magnetometry based on nitrogen-vacancy (NV) centers in nanodiamonds was employed to probe the magnetic domains on the sample surface. As shown in Figure 4a, nanodiamonds with ensemble NV centers were mounted on a cantilever probe and continuous-wave optically detected magnetic resonance (ODMR) spectra were measured by varying the tip position relative to the sample surface under zero magnetic field, to investigate the stray fields of the sample. Figure 4b and 4c show the typical ODMR spectra from NV tip during scanning the $Mn_3Ga$ alloy and $Mn_{1-x}Ga_xN$ films, respectively. The resonant peak splitting observed in the $Mn_3Ga$ alloy film, attributed to the local stray magnetic field (Zeeman effect), disappears in $Mn_{1-x}Ga_xN$ films after annealing, indicating that structural modifications and nitrogen incorporation during annealing lead to the loss of magnetism. In order to investigate the changes in the magnetic domain structure of the film, mapping measurements were conducted over a 1.6 μm area using a diamond probe positioned at a height of 150 nm. The corresponding results are presented in Figures 4d and 4e. The ODMR spectra of the Mn₃Ga films exhibit significant peak splitting, indicating the presence of well-defined magnetic domain structures (Figure S4, Supporting Information). In contrast, no peak splitting was observed in the $Mn_{1-x}Ga_xN$ film within the measured regions (Figure S4, Supporting Information), further confirming the macroscopic magnetic changes in the sample at a larger scale. These results had been repeatedly observed in several random locations, further elaborating the robustness of our results.

**Changes in transport behavior of manganese-based films**

To evaluate the impact of the ammonia annealing process on the electrical properties of the samples, we conducted electrical transport measurements on Mn₃Ga alloy and $Mn_{1-x}Ga_xN$ films. Figure 5a shows the temperature dependent resistivity (ρ-T) curves, both samples exhibit metallic behavior across all temperatures, with relatively low resistivity values ranging from $10^{-4}$ to $10^{-5}$ Ω·cm. Compared to the alloy thin films, the resistivity of the nitrogenated samples shows a more gradual temperature dependence. Magnetoresistances [MR, $(\rho_{xx}-\rho_0)/\rho_0$] of specimens were recorded by applying currents along the *a* and *b* directions at various temperature (Figure S5, Supporting Information). The inset of Figure 5a shows MRs of Mn₃Ga alloy and $Mn_{1-x}Ga_xN$ films at 10 K when *H//c*. We noticed that anomalous butterfly-shaped magnetoresistance loop vanishes after annealing, with the MR of the $Mn_{1-x}Ga_xN$ film remaining consistently close to zero. Furthermore, Hall resistivity ($\rho_{xy}$) were also conducted when *H//c* to probe the intriguing magnetic properties of the Mn₃Ga alloy and $Mn_{1-x}Ga_xN$ films. In ferromagnetic materials, the Hall resistivity ($\rho_{xy}$) is typically expressed as $\rho_{xy}$ = R₀H + R_SM_Z, where R₀ represents the ordinary Hall effect (OHE) coefficient, R_S denotes the anomalous Hall effect (AHE) coefficient, and M_Z is the out-of-plane magnetization. The ordinary Hall resistivity (R₀H) was subtracted from $\rho_{xy}$ to separate the anomalous Hall effect (AHE) contribution from Hall resistivity. Figure 5b presents the field-dependent ($\rho_{xy}$ − R₀H) at various temperatures for Mn₃Ga alloy films, as well as $\rho_{xy}$ for $Mn_{1-x}Ga_xN$ films. At 3 K, $\rho_{xy}$ − R₀H for Mn₃Ga exhibits a square-like hysteresis loop, while $\rho_{xy}$ − R₀H for $Mn_{1-x}Ga_xN$ shows negligible values. The saturation ($\rho_{xy}$ − $R_0H$) exhibits a positive correlation with temperature, increasing as the temperature rises (Figure 5c). At temperatures of 50 K and above, the anomalous Hall resistivity comprises two overlapping hysteresis loops, attributed to the two magnetic contributions from polycrystalline structure of the $Mn_3Ga$ alloy thin films. Additionally, the Hall coefficient ($R_H$, Figure 5d) and carrier density ($n$, Figure 5e) of $Mn_3Ga$ alloy and $Mn_{1-x}Ga_xN$ thin films were measured across various temperature. $R_H$ of $Mn_{1-x}Ga_xN$ is approximately an order of magnitude lower than that of $Mn_3Ga$ and decreases with increasing temperature, in contrast to the trend observed in $Mn_3Ga$. For $R_H = \frac{1}{ne}$, the carrier density ($n$) inversely correlates with the Hall coefficient. $Mn_{1-x}Ga_xN$ exhibits a high carrier concentration of approximately $(1-1.5) \times 10^{23}$ $cm^{-3}$. Therefore, the transport measurements confirm the significant changes in the electronic ground states of $Mn_{1-x}Ga_xN$ after nitridation.

**Conclusions**

This study presents a novel and effective strategy for synthesizing high-quality single-crystal nitride thin films via high-temperature ammoniation of alloy thin films. X-ray diffraction analysis confirms the excellent epitaxial quality of the nitrided samples, while nitrogen incorporation into the lattice induces significant structural and magnetic transformations. The formation of nitrogen-metal bonds alters orbital occupancy, shifting from out-of-plane to in-plane states under an external magnetic field, consistent with the antiferromagnetic character. Diamond NV center magnetometry reveals a zero net magnetic moment in nitride thin films, signifying the suppression of ferromagnetism in alloy films, a finding further supported by electrical transport measurements. This work establishes a robust framework for the fabrication of complex nitride materials, offering a promising avenue for next-generation spintronic devices. Beyond the fundamental insights gained, this method provides a pathway for engineering novel antiferromagnetic materials with tunable electronic and magnetic properties.

**Experimental Section**

Thin film synthesis

The nitride thin films were prepared in two steps. First, $Mn_3Ga$ alloy thin films were grown on (001)-oriented MgO substrates via pulsed laser deposition (PLD) using a stoichiometric

Mn₃Ga alloy target (99.9% purity, Beijing Loyaltarget Tech. Co., LTD). During deposition, the substrate temperature was maintained at 500°C, and the base pressure was kept at $1 \times 10^{-8}$ Torr. The laser energy density and repetition rate were set at 1.25 J cm$^{-2}$ and 5 Hz, respectively. Following deposition, the Mn₃Ga alloy films were cooled to room temperature and subjected to rapid thermal annealing (RTA). The annealing process involved heating the films to 700°C within 5 minutes under a pure ammonia atmosphere, maintaining this temperature for 1 hour, and then allowing the films to cool slowly to room temperature.

Structural characterizations

X-ray diffraction (XRD) θ–2θ scans, X-ray reflectivity (XRR), and reciprocal space mapping (RSM) were performed using a Panalytical X'Pert³ MRD diffractometer with Cu Kα₁ radiation and a 3D pixel detector. Sample thicknesses were determined by fitting XRR curves using GenX software. Cross-sectional scanning transmission electron microscopy (STEM) samples were prepared via mechanical thinning followed by ion milling. High-angle annular dark-field (HAADF) and annular bright-field (ABF) imaging were conducted in scanning mode using a JEM ARM 200CF microscope at the Institute of Physics (IOP), Chinese Academy of Sciences (CAS). Optical SHG measurements on Mn₃Ga alloy and Mn₃GaN thin films were carried out at University of Electronic Science and Technology of China. All measurements were performed were performed on independent home-developed systems at room temperature. The pumping beam utilized is a femtosecond fiber laser (Femto-IR-200-40, Huaray Precision Laser, λ=800nm). Linearly polarized light is directed onto the sample at an incidence angle of 45°. The polarization direction of the incident light was precisely adjusted using a half-wave plate (λ/2, 400nm). Subsequently, the polarization of the reflected light was aligned to be parallel to that of the incident beam ($I_{p-out}^{2\omega}$). The resulting optical signals were then captured and measured using a photomultiplier tube, ensuring accurate detection.

Spectroscopic measurements

Elemental-specific X-ray absorption spectra (XAS) measurements were conducted at room temperature in the total electron yield (TEY) mode by collecting the sample-to-ground drain current at both N *K*-edges and Cr *L*-edges on the MCD beamline of National Synchrotron Radiation Laboratory (NSRL) in Hefei, China. The XLD measurements were performed by changing incidence angle (α) of linearly polarized X-ray beam with respect to the sample's surface plane. When X-ray scattering plane was rotated to 0° (α = 0°), that is, the X-ray beam was parallel to the surface plane, the XAS signal directly reflects the $d_{x^2-y^2}$ orbital occupancy. While the angle change to 60° (α = 60°), XAS probes the orbital information from both $d_{x^2-y^2}$ and $d_{3z^2-r^2}$ orbitals. For simplifying the calculations, the XLD signals of $Mn_{1-x}Ga_xN$ films with zero and 0.4T magnetic field were calculated by $I_{0°}$ - $I_{60°}$. Instead of switching the polarity of X-ray beams, we collected the XAS under the opposite magnetic fields with amplitude of 0.4 T. X-ray photoelectron spectroscopy (XPS) measurements were performed at the Institute of Physics (IOP), Chinese Academy of Sciences (CAS). Spectra were collected at Mn 2p, Ga 3d and N 1s core-level peaks at room-temperature, respectively.

Diamond NV-based magnetometry

Diamond nitrogen-vacancy (NV) center-based magnetometry measurements were performed at room temperature using a home-built optically detected magnetic resonance (ODMR) system. The NV center was mounted on the tip of a cantilever probe, while the sample was affixed to motorized stages capable of precise micrometer-scale movement, enabling spatial mapping of the magnetic signal on the sample surface. At zero magnetic field, the $m_S = \pm 1$ states of the NV center are degenerate, resulting in a single resonance dip at around 2870 MHz in the ODMR spectrum. When a stray magnetic field is present, the transition frequencies of $m_S = +1$ and $m_S = -1$ to $m_S = 0$ shift linearly due to the Zeeman effect, causing the ODMR spectrum to split into two distinct resonance dips.

Electrical transport and magnetization measurements

Electrical transport measurements were performed using the standard van der Pauw geometry with aluminum contacts in a Physical Property Measurement System (PPMS). Resistivity and Hall conductivity were measured under an applied magnetic field of $\mu_0 H = \pm 5$ T. Macroscopic magnetization measurements were conducted using a Magnetic Property Measurement System (MPMS) with in-plane magnetic fields. *M-H* hysteresis loops were recorded at room temperature, while *M-T* curves were obtained during field-cooled (FC) and zero-field-cooled (ZFC) processes after magnetizing the sample with a 1 kOe field.

**Acknowledgements**

This work was supported by the National Key Basic Research Program of China (Grant Nos. 2020YFA0309100 to E.J.G., 2022YFA1403000 to J.Q.), the Beijing Natural Science Foundation (Grant No. JQ24002 to E.J.G.), the National Natural Science Foundation of China (Grant Nos. U22A20263 to E.J.G., 52250308 to E.J.G., 12304158 to Q.J., 12434003 J.Q. and 12474096 to C.W.), , the CAS Project for Young Scientists in Basic Research (Grant No. YSBR-084 to E.J.G.), the CAS Youth Interdisciplinary Team, the Special Research assistant, the CAS Strategic Priority Research Program (B) (Grant No. XDB33030200 to K.J.J. and E.J.G.), the Guangdong Basic and Applied Basic Research Foundation (Grant No. 2022B1515120014 to E.J.G.), the Guangdong-Hong Kong-Macao Joint Laboratory for Neutron Scattering Science and Technology, the China Postdoctoral Science Foundation (Grant No. 2022M723353 to E.J.G.), and the International Young Scientist Fellowship of IOP-CAS to S.C. XPS experiments were performed at IOP-CAS via a user proposal. The synchrotron-based XAS and XLD was conducted at the National Synchrotron Radiation Laboratory, Hefei.

**Conflict of Interest**

The authors declare no conflict of interest.

**Data Availability Statement**

The datasets generated during and/or analyses during the current study are available from the first author (Q.Y.W.) and corresponding authors (E.J.G.) on reasonable request.

**Figure 1**

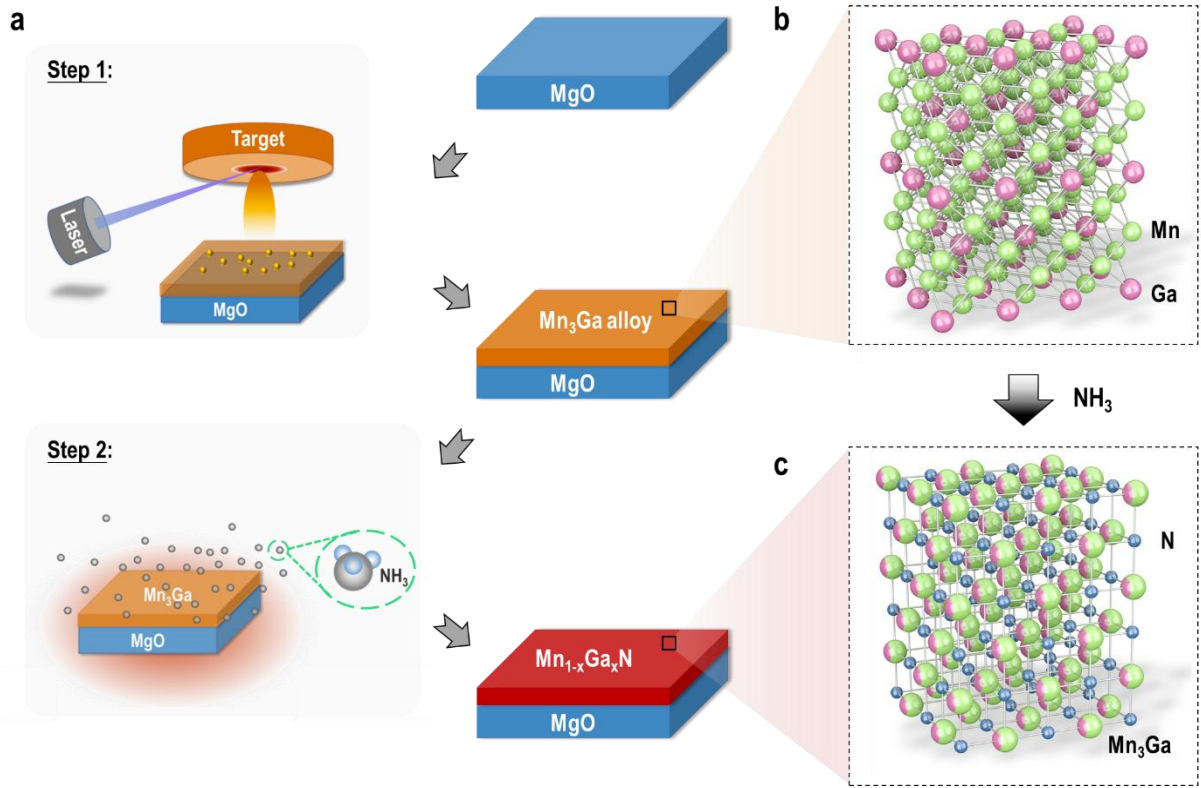

**Figure 1. Synthesis of Mn$_{1-x}$Ga$_x$N (x~0.25) thin films.** (a) Schematic of the fabrication process. Mn$_3$Ga alloy thin films were firstly deposited on (001)-oriented MgO substrates using pulsed laser deposition (PLD). The films were then subjected to rapid thermal annealing (RTP) in an ammonia atmosphere at a partial pressure of 10$^4$ Pa and a temperature of 700°C, inducing nitrogen incorporation and crystallization. (b) and (c) Schematic representations of the crystal structures of Mn$_3$Ga and Mn$_{1-x}$Ga$_x$N, respectively.

**Figure 2**

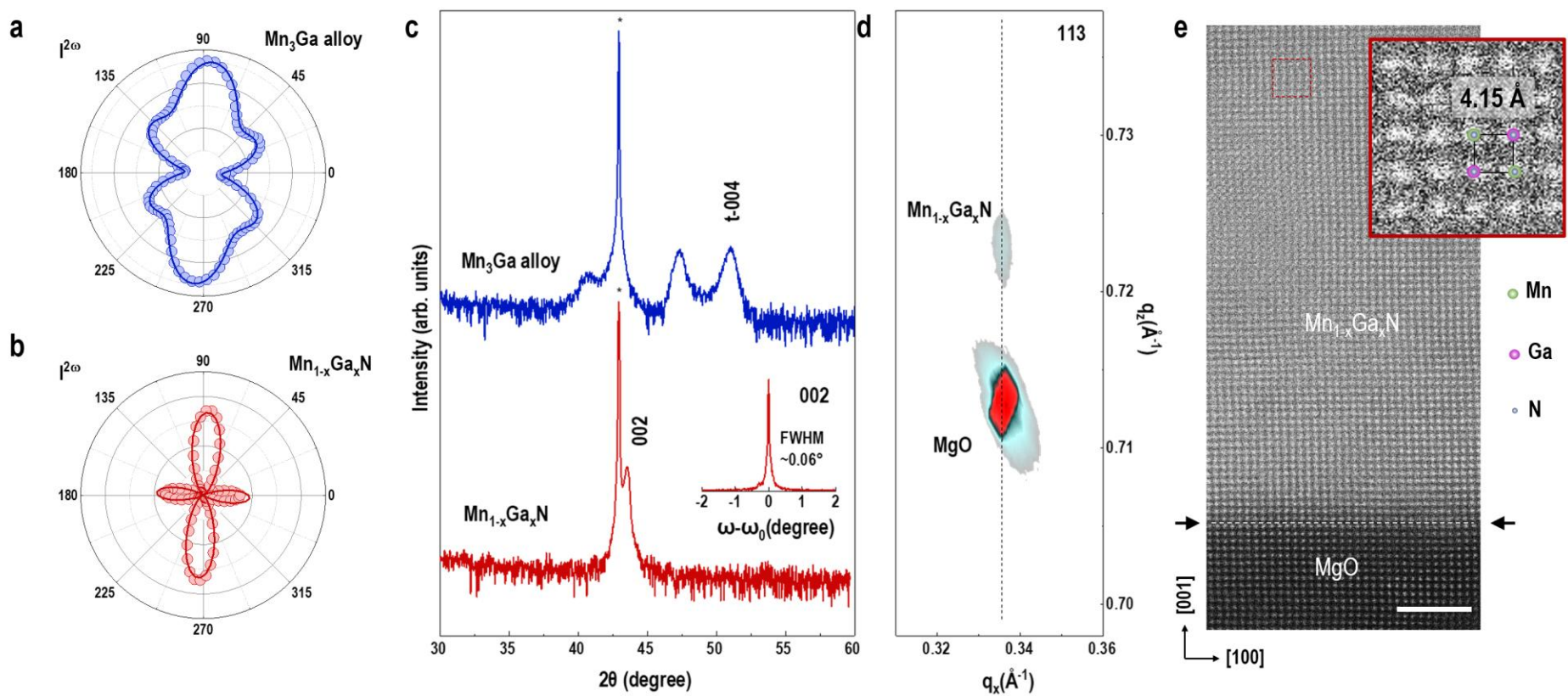

**Figure 2. Structural transition induced by ammonia nitriding.** (a) and (b) SHG signals (scatter points) and theoretical fits (solid lines) for $Mn_3Ga$ alloy and $Mn_{1-x}Ga_xN$ (x~0.25) thin films, respectively. The analyzer polarization for SHG signal (400 nm) is parallel to that of the incident beam (800 nm), while the sample is rotated. (c) XRD θ–2θ scans of $Mn_3Ga$ alloy and $Mn_{1-x}Ga_xN$ thin films. The inset shows the rocking curve of the $Mn_{1-x}Ga_xN$ (002) reflection, with a FWHM of ≈0.06°. (d) RSM of a $Mn_{1-x}Ga_xN$ thin film taken around the (113) reflection of the MgO substrate. (e) Cross-sectional HAADF-STEM image of the $Mn_{1-x}Ga_xN$ /MgO heterointerface, confirming the *c*-axis orientation and coherent growth of the single-crystalline thin film. The inset presents a zoomed-in STEM image of the $Mn_{1-x}Ga_xN$ film, corresponding to the red dashed square.

<i>10</i> **Figure 3**

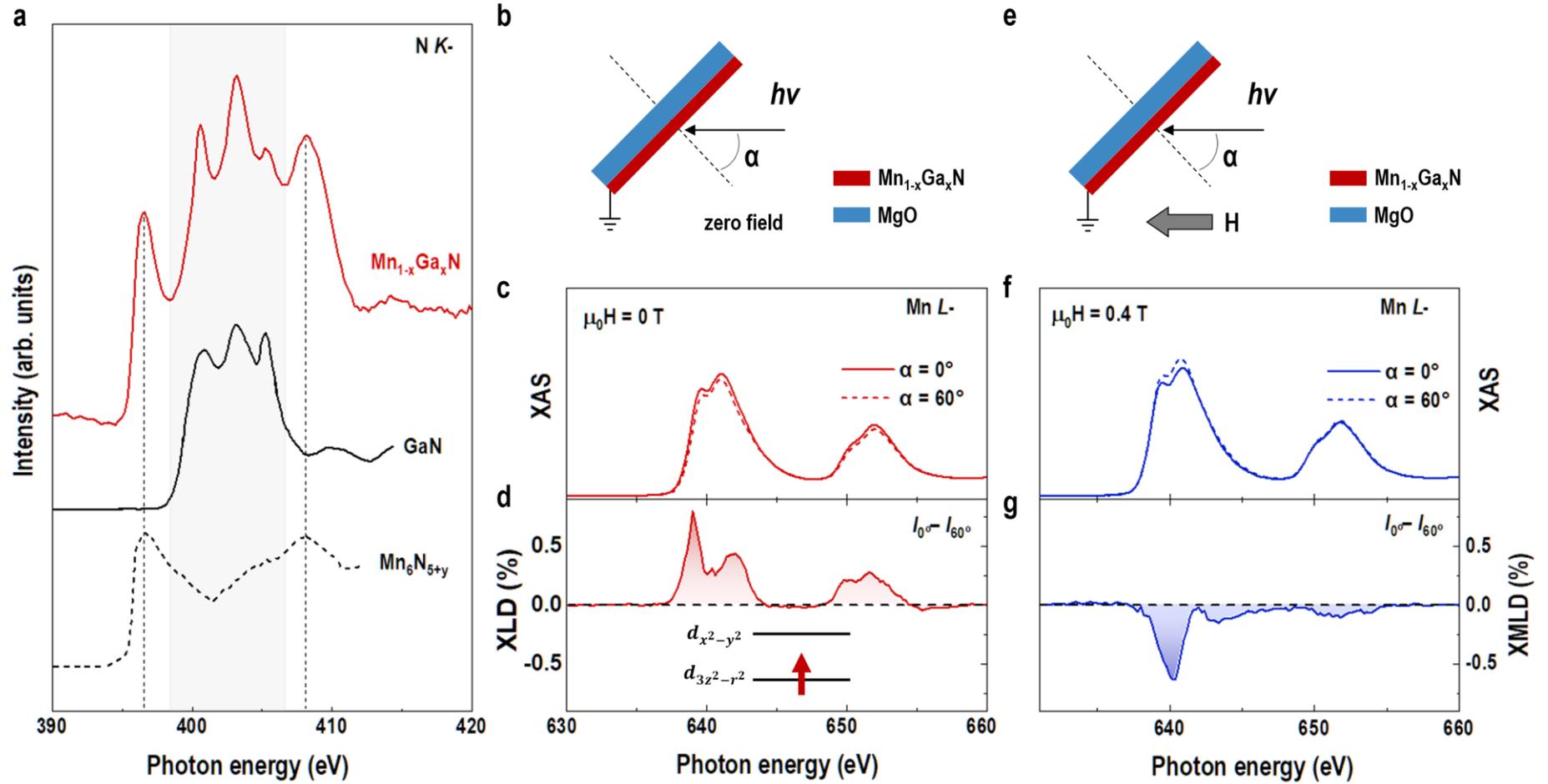

**Figure 3. Soft XAS of Mn$_{1-x}$Ga$_x$N (x~0.25) thin films.** (a) XAS at the N $K$-edges for Mn$_{1-x}$Ga$_x$N thin films. Dashed lines represent reference N $K$-edge spectra for GaN and θ-Mn$_6$N$_{5+y}$ (y = 0.26). The gray-shaded area highlights the peak positions corresponding to N–Mn and N–Ga bond states. (b) and (e) Schematics of the measurement setups under zero field and an applied magnetic field, respectively. (c) and (f) XAS at the Mn

15  *L*-edges for $Mn_{1-x}Ga_xN$ thin films measured under zero field and a 0.4 T magnetic field, respectively. The sample's scattering plane was rotated at angles of 0° and 60° with respect to the incident x-ray beam direction. (d) and (g) X-ray linear dichroism (XLD) and X-ray magnetic linear dichroism (XMLD) spectra, calculated from the intensity difference between $I_{0°}$ and $I_{60°}$ under zero field and a 0.4 T magnetic field, respectively. The clear XLD and XMLD signals indicate a magnetic field-dependent electron occupancy in the Mn $3d$ $e_g$ orbitals.

**Figure 4**

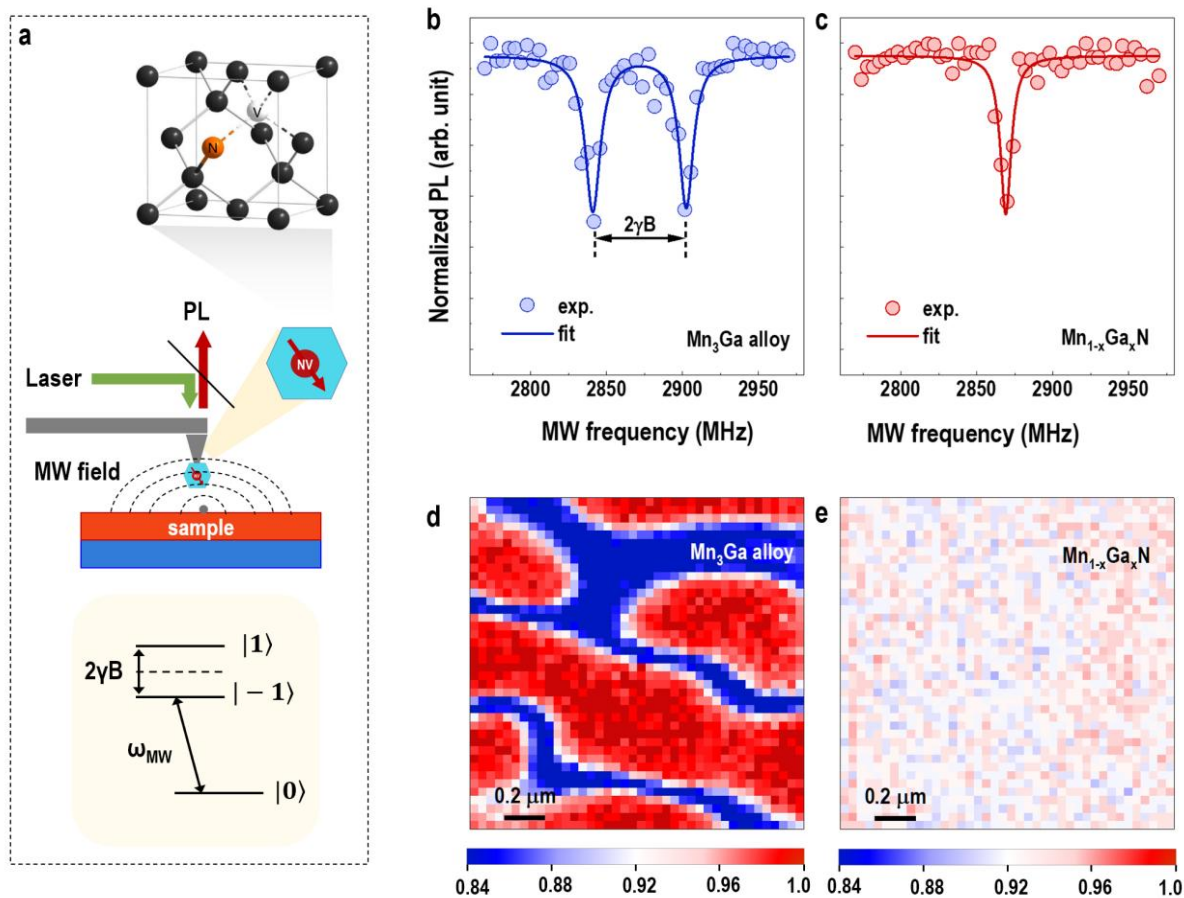

**Figure 4. Nanoscale magnetic probing via scanning NV magnetometry.** (a) Schematic of diamond NV magnetometry, where an NV center in a diamond tip is mounted on a cantilever probe and scanned to detect photoluminescence spectra at different locations. (b) and (c) Optically detected magnetic resonance (ODMR) spectra of ensemble NV centers in nanodiamonds on $Mn_3Ga$ alloy and $Mn_{1-x}Ga_xN$ thin films, respectively. (d) and (e) Zero-field ODMR mappings of $Mn_3Ga$ alloy and $Mn_{1-x}Ga_xN$ thin films, respectively. The *x* and *y* axes span 1.6 μm, while the *z* position is fixed at a height of 150 nm. The significant contrast in $Mn_3Ga$ alloy films indicates ferromagnetism, whereas $Mn_{1-x}Ga_xN$ films do not exhibit a clear magnetic signal.

**Figure 5**

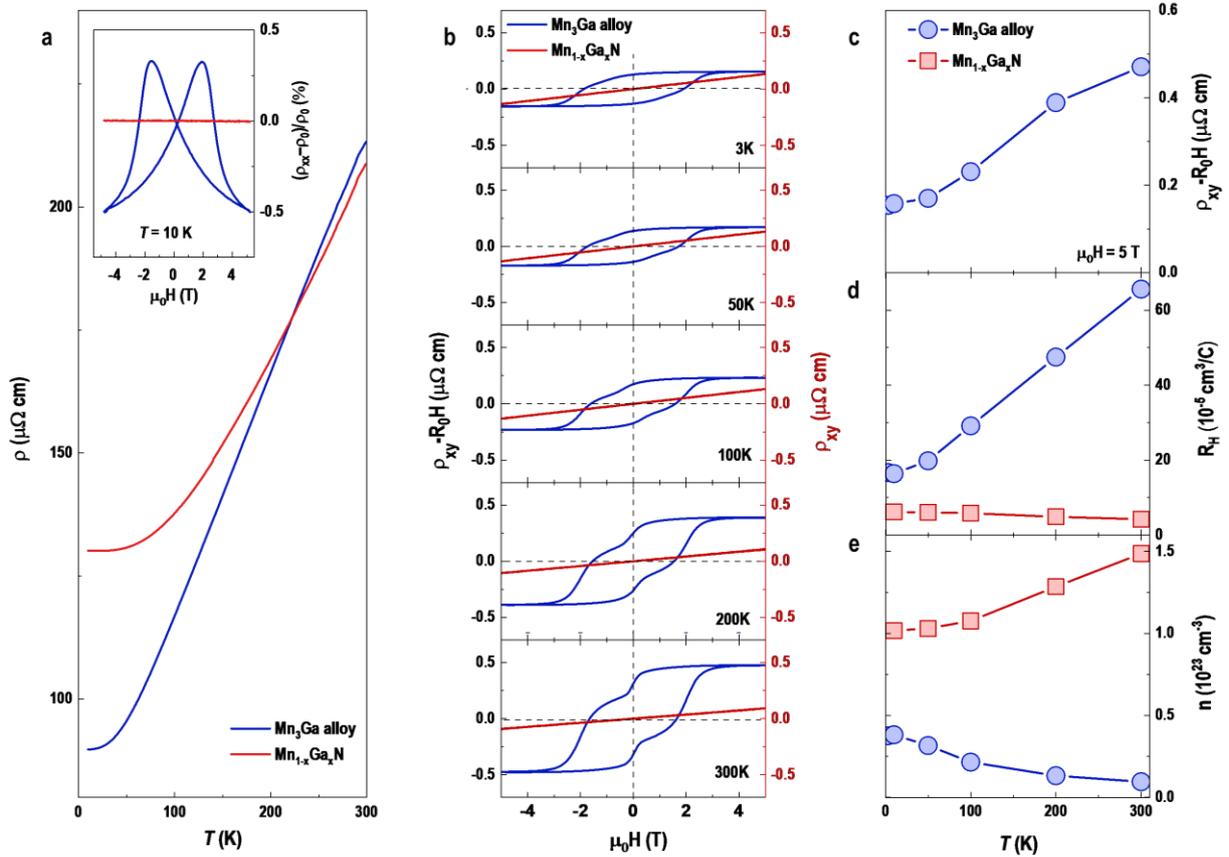

**Figure 5. Transport behaviors of Mn$_3$Ga alloy and Mn$_{1-x}$Ga$_x$N (x~0.25) films.** (a) Temperature-dependent resistivity ($\rho$–T) curves of Mn$_3$Ga and Mn$_{1-x}$Ga$_x$N films. The inset shows the magnetoresistance ($\rho_{xx}$–$\rho_0$)/$\rho_0$ as a function of magnetic field at 10 K. (b) Field-dependent anomalous Hall resistivity ($\rho_{xy}$–$R_0H$) of Mn$_3$Ga alloy and Mn$_{1-x}$Ga$_x$N thin films at different temperatures. (c) $\rho_{xy}$–$R_0H$ of Mn$_3$Ga alloy thin films at 5 T, (d) Hall coefficient ($R_H$), and (e) carrier density ($n$) of Mn$_3$Ga alloy and Mn$_{1-x}$Ga$_x$N thin films as a function of temperature.